\documentclass[epj]{svjour}
\usepackage{latexsym}
\usepackage{graphics}
\usepackage{graphicx}
\usepackage{dcolumn}
\usepackage{booktabs, array}
\usepackage{siunitx}
\usepackage{supertabular}
\usepackage{multirow}
\usepackage{booktabs}
\usepackage{bm}
\usepackage{sidecap}
\usepackage{ulem}
\usepackage{hyperref}
\usepackage{cite}
\usepackage{siunitx}
\usepackage{lipsum}
\usepackage{microtype}
\usepackage{amssymb}
\usepackage[utf8]{inputenc}
\usepackage[english]{babel}
\usepackage{amsmath}
\usepackage{longtable}
\usepackage{geometry}
\usepackage{float}
\usepackage{placeins}

\begin{document}\sloppy

%
\title{Cluster radioactivity from trans-tin to superheavy region using an improved empirical formula}
\author{G. Saxena\inst{1,2} \and A. Jain \inst{3,4}
}                     
%
%
\institute{Department of Physics (H\&S), Govt. Women Engineering College, Ajmer-305002, India \and Department of Physics, Faculty of Science, University of Zagreb, Bijeni$\breve{c}$ka c. 32, 10000 Zagreb, Croatia \and Department of Physics, School of Basic Sciences, Manipal University Jaipur, Jaipur-303007, India \and Department of Physics, S. S. Jain Subodh P.G.(Autonomous) College, Jaipur-302004, India}
\date{Received: date / Revised version: date}
\abstract{A simple relation $(aZ_{c} + b)(Z_{d}/Q)^{1/2} + (cZ_{c} + d)$ of estimation of the half-life of cluster emission is further improved for cluster and $\alpha$-decays, separately, by incorporating isospin of parent nucleus as well as angular momentum taken away by the emitted particle. This improved version is not only found robust in producing experimental half-lives belonging to the trans-tin and trans-lead regions but also elucidates cluster emission in superheavy nuclei over the usual $\alpha$-decay. Considering daughter nuclei around the doubly magic $^{100}$Sn and $^{208}$Pb nuclei for trans-tin and trans-lead (including superheavy) parents, respectively, a systematic and extensive study of 56$\leq$Z$\leq$120 isotopes is performed for the light and heavy cluster emissions. A fair competition among cluster emission, $\alpha$-decay, spontaneous fission, and $\beta$-decay is observed in this broad range resulting in a substantial probability of C to Sr clusters from several nuclei, which demonstrates the adequacy of shell effects. The present article proposes a single, improved, latest-fitted, and effective formula of cluster radioactivity that can be used to estimate precise half-lives for a wide range of the periodic chart from trans-tin to superheavy nuclei.
} 
\authorrunning{G. Saxena and A. Jain}
\titlerunning{Cluster radioactivity from trans-tin to superheavy region....}
\maketitle
\section{Introduction}
\label{intro}
An exotic decay mode, in which a heavy nucleus fragmentizes itself by emitting a nucleus with the mass between the fission product and $\alpha$-particle, is envisioned already in 1980 \cite{sandulescu1980} and named 'cluster decay'. After its first experimental confirmation \cite{rose1984} by emission of $^{14}$C from actinide parent nucleus $^{223}$Ra, this decay mode has been observed from many trans-lead nuclei as the emission of light $^{14}$C to heavy $^{32}$Si clusters \cite{Bonetti2007}. From the last two decades it has gained healthy attention \cite{Kumar2003,Kumar2009,Gupta2003,Gupta1994,Price1989,Santhosh2012,Hourani1989,Royer2001,Singh2022,kulkin2005} due to the direct signature of the crucial role of shell effects as all known cluster emissions have daughter nuclei near to the closed-shell nucleus $^{208}$Pb.\par

Except for the trans-lead region, there are two potential and new islands of cluster emitters that are still less explored and require a comprehensive investigation. The first one is the trans-tin region, regulated by doubly magic nucleus $^{100}$Sn, where only one experimental observation of $^{12}$C cluster decay from $^{114}$Ba exists \cite{Og1994,Guglielmetti1995,Guglielmetti1995npa}, though it is not observed in the later measurement \cite{Guglielmetti1995}. The other region is the superheavy region in which many heavy clusters are speculated to compete with $\alpha$-decay \cite{poenaru2012,zhang2018,poenaru2018,poenaru2018EPJA,matheson2018,Warda2018,santhosh2008,prathapan2021,Santhosh2022prc,Royer2022,Nithya2022}, however, still requires experimental confirmation \cite{santhosh2018,soylu2019,jain2023PS}. Therefore, an exhaustive study of cluster radioactivity including trans-tin, trans-lead as well as superheavy regions can lead to important structure information for the detection of heavy and superheavy nuclei in future experimental investigations.\par

The estimation of the half-life of cluster decay was made in the same decade of its experimental confirmation by several theoretical models which are primarily based on the quantum tunneling effect through a potential barrier \cite{Poenaru1982,Delion2001,Poenaru1996,Poenaru1996IPPB,arun2009} or cold asymmetric fission process \cite{Brack1972,Wahl1988}. Alternatively and concomitantly, the very first estimation of the half-life of heavy ion (cluster) emission was given by Poenaru \textit{et al.} using an analytical formula based on an analytical superheavy asymmetric fission model (ASAFM) \cite{Poenaru1985,Poenaru1986}. Since then, cluster radioactivity has been studied by several models out of which a few recent ones are the preformed-cluster-decay model (PCM) \cite{Joshua2022,Yahya2022}, Coulomb and proximity potential model for deformed nuclei \cite{Hosseini2023}, Wentzel-Kramers-Brilliouin (WKB) method \cite{Dagtas2022}, and modified Gamow-like model \cite{Liu2021}. Similarly, by fitting available experimental data, many (semi)empirical/analytical formulas related to cluster decay were proposed from time to time such as Horoi formula \cite{Horoi2004}, BKAG formula \cite{Balasubramaniam2004}, RenA formula \cite{Ren2004}, NRDX formula \cite{NRDX2008}, UDL formula \cite{UDL2009}, and UNIV formula \cite{Poenaru2011}. Later on, Tavares and Medeiros proposed a semiempirical relationship for the accurate estimation of exotic decay from trans-lead nuclei \cite{Tavares2013}. Recently, a few of above mentioned  semiempirical relationships are modified in Refs. \cite{jain2023,soylu2021,UF2022,Wang2021CPC,cheng2022,Lin2023} which endorse the crucial contributions of asymmetry of parent nucleus and centrifugal barrier effect in the estimation of half-lives of cluster radioactivity. It is, however, important to point out here that almost all the formulas (older or modified) work fairly well in the trans-lead region, but act somewhat unfavourable for the simultaneous estimation of half-lives of cluster emitters in trans-tin and superheavy regions \cite{nature,zhang2018}. As a matter of fact, especially for the trans-tin region, the UDL formula overestimates the half-life of $^{114}$Ba by four orders of magnitude whereas it works well in the superheavy region due to its fission-like mechanisms \cite{zhang2018}. On the other hand, Horoi and NRDX formulas are capable to reproduce the half-life of $^{114}$Ba \cite{nature,Guglielmetti1997PRC} but are found not suitable for the heavy clusters in the superheavy region \cite{zhang2018}. Therefore, there is a need for a unique formula that can be applied for the accurate estimation of cluster decay half-life emulated with $\alpha$-decay half-life for the whole periodic chart (from trans-tin to superheavy nuclei), which is precisely the aim of the present investigation. \par

In this article, we present a single formula that can be applied for the accurate estimation of half-lives of cluster emission for a systematic study of nuclei in the range 56$\leq$Z$\leq$120 considering daughter nuclei around the doubly magic nuclei $^{100}$Sn and $^{208}$Pb for the trans-tin and trans-lead (including superheavy region) regions, respectively. The most probable clusters are found as a result of its contest with $\alpha$-decay, $\beta$-decay (referring for $\beta^{-}$, $\beta^{+}$, and EC throughout the paper), and spontaneous fission (SF) for this wide region of the periodic chart.\par
\section{Formalism}
\begin{table*}[!htbp]
\caption{The coefficients of ITM formula proposed in the present work. The formula is fitted separately for trans-tin and tans-lead regions (see the text for details).}
\centering
\def\arraystretch{1.0}
\resizebox{1.0\textwidth}{!}{%
{\begin{tabular}{c|cc|cc}

 \hline
\multicolumn{1}{c|}{Coefficient}&
\multicolumn{2}{c|}{Trans-tin}&
\multicolumn{2}{c}{Trans-lead}\\
\cline{2-5}
\multicolumn{1}{c|}{}&
 \multicolumn{1}{c}{ITM-CT}&
\multicolumn{1}{c|}{ITM-AT}&
\multicolumn{1}{c}{ITM-CL}&
\multicolumn{1}{c}{ITM-AL}\\
 \hline
 a&14.5427$\pm$0.8701    & 294.8314$\pm$52.3194   & 11.4570$\pm$0.0185  & 217.4129$\pm$42.6150   \\
 b&-1.5239$\pm$0.5639    & -578.2390$\mp$104.6389  & -8.9167$\mp$0.0446 &-420.0786$\mp$ 85.2188   \\
 c&-3.9917$\mp$0.0639    & -10.4338$\mp$0.1024    & -4.2778 $\mp$0.0131 & -12.5941$\mp$ 0.1005    \\
 d&-109.0040$\mp$10.4162 & -22.8685$\mp$0.1952   & -103.9499$\mp$0.2137& -27.1875$\mp$ 0.2003    \\
 e&-10.8146$\mp$3.6985  & 11.8588$\mp$1E-6      & 93.2694$\mp$0.4313  & 4.2401$\mp$0.0845    \\
 f&0.6419$\pm$0.1954 & 0.1495$\mp$1E-6      & 0.3527$\pm$0.0009  & 0.2630$\mp$0.0009    \\
\hline
\end{tabular}}}
\label{tab:coefficient}
\end{table*}

In 2013, Tavares and Medeiros found that experimental cluster decay half-lives show straight lines when plotted as a function of $\xi$ = $(Z_{d}/Q)^{1/2}$, where $Z_{d}$ is the atomic number of daughter and $Q$ is the energy of the two-body disintegrating system \cite{Tavares2013}. A formula (TM), shown below,  was proposed for cluster decay half-lives dependent on only 4 parameters and was able to reproduce the majority of the available experimental half-life data reasonably well which belong to the trans-lead region.
\begin{eqnarray}
log_{10}T_{1/2}^{TM}(s) = (aZ_{c} + b) \sqrt{\frac{Z_{d}}{Q}} + (cZ_{c} + d),
    \label{eqtm}
    \end{eqnarray}
In our investigation, this relationship while refitted, is also found to reproduce $\alpha$-decay half-lives of the latest evaluated database NUBASE2020 \cite{audi20201} with a good accuracy for the wide range of the periodic chart. Similar to the crucial roles of (i) angular momentum ($l$) taken away by the emitted particle, and (ii) isospin ($I = (N-Z)/A$) of parent nucleus in $\alpha$-emission \cite{sharma2021npa}, these two quantities are reported equally important in determining the cluster decay half-lives by new UDL formula of Soylu and Qi \cite{soylu2021} as well as by improved NRDX formula (named as improved unified formula (IUF)) of Ismail \textit{et al.} \cite{UF2022}. In addition, the crucialness of the isospin effect is also probed by improving the semi-empirical formula (ISEM) for the cluster radioactivity in Ref. \cite{cheng2022}. Recently, a few semiempirical relationships are modified for the trans-lead region in Ref. \cite{jain2023} which endorse the crucial contributions of asymmetry of parent nucleus and centrifugal barrier effect in estimating half-lives of cluster radioactivity. In order to elucidate the effect of these terms in Eqn. (\ref{eqtm}), we have initially fitted our data set for (i) cluster decay and (ii) $\alpha$-decay in the trans-tin region together with (iii) cluster decay and (iv) $\alpha$-decay in the trans-lead region by incorporating only the $I$-dependent term ($\sqrt{I(I+1)}$), which accounts for the asymmetry. This addition of term has yielded a reduction of RMSE values from 1.53, 1.21, 0.97, and 0.92 to 1.37, 0.74, 0.84, and 0.89, respectively. Subsequently, when we have introduced only the $l$-dependent term ($\sqrt{l(l+1)}$), which reflects the hindrance effect of the centrifugal barrier, then the RMSE values reduce to 1.01, 1.09, 0.75, and 0.78, respectively. Finally, when both the $I$ and $l$ dependent terms are incorporated in Eqn. (\ref{eqtm}) then RMSE values significantly reduced to 0.82, 0.71, 0.62, and 0.76, respectively.  Hence, the Eqn. (\ref{eqtm}) can be improvised by adding isospin dependent term as well as a centrifugal barrier term which lead to the following improved formula (named as improved Tavares and Medeiros formula or ITM formula):\\
    \begin{eqnarray}
     log_{10}T_{1/2}^{ITM}(s) &=& (aZ_{c} + b) \sqrt{\frac{Z_{d}}{Q}} + (cZ_{c} + d)  \nonumber \\ &&+e \sqrt{I(I+1)}
    \nonumber \\ &&+ f \sqrt{l(l+1)},
    \label{eqmtm}
    \end{eqnarray}

where, a, b, c, d, e, and f are fitting coefficients, tabulated in Table \ref{tab:coefficient}. Z$_{c}$ and Z$_{d}$ are the proton number of emitted cluster and daughter nucleus, respectively. $Q$ and $I$ are the disintegration energy and isospin, respectively. $l$ is the minimum angular momentum of the cluster particle, which is obtained by following selection rules:
\begin{equation}
   l=\left\{
    \begin{array}{ll}
       \triangle_j\,\,\,\,\,\,
       &\mbox{for even}\,\,\triangle_j\,\,\mbox{and}\,\,\pi_{i} = \pi_{f}\\
       \triangle_{j}+1\,\,\,\,\,\,
       &\mbox{for even}\,\,\triangle_j\,\,\mbox{and}\,\,\pi_{i} \neq \pi_{f}\\
       \triangle_{j}\,\,\,\,\,\,
       &\mbox{for odd}\,\,\triangle_j\,\,\mbox{and}\,\,\pi_{i} \neq \pi_{f}\\
       \triangle_{j}+1\,\,\,\,\,\,
       &\mbox{for odd}\,\,\triangle_j\,\,\mbox{and}\,\,\pi_{i} = \pi_{f},\\
      \end{array}\right.
      \label{lmin}
\end{equation}
here, $\triangle_j$ = $|j_p - j_d - j_c|$ with j$_{p}$, $\pi_{i}$, are the spin and parity values of the parent nucleus, respectively. j$_{d}$ is the spin of the daughter nucleus. $\pi_{f} = (\pi_{d})(\pi_{c})$, in which, $\pi_{d}$ and $\pi_{c}$ are the parities of the daughter nucleus and cluster, respectively. For the purpose of fitting, the data of spin and parity are taken from NUBASE2020 \cite{audi20201}.\par

Before discussing the results obtained by the ITM formula, there are two important facts that are worth mentioning here. The first fact is related to the inclusion of deformation which is an important ingredient for the estimation of half-lives as also has been described in our recent article \cite{saxenajpg2023}. We have followed a similar analysis by incorporating the deformation of the parent nucleus but the reduction in RMSE values is found very insignificant on the cost of the addition of one more parameter. In several articles \cite{Soylu2012epja,Hosseini2023}, it is demonstrated that in addition to the deformation of the parent nucleus, the deformation of cluster and daughter nuclei needs to be considered to visualize the complete picture of this exotic decay. Such kind of investigation requires a comprehensive and detailed analysis with the inclusion of quadrupole $(\beta_2)$ and hexadecapole $(\beta_4)$ deformations, which will be reported in our subsequent work on the deformation effect of $\alpha$ and cluster decays along with the effect of shape-coexistence on half-lives. However, when compared to the half-lives of cluster emission considering the deformation effect from Refs. \cite{Soylu2012epja,Hosseini2023}, the present form of the ITM formula results reasonably well with the RMSE value of 0.62 instead of RMSE values 2.30 of Ref. \cite{Soylu2012epja} and 0.74 of Ref. \cite{Hosseini2023} on the provided experimental data.\par

Another fact is related to the internal structure of the heavy cluster. For simplicity, in general, cluster formation is assumed similar to the formation of $\alpha$-particle inside the nucleus but this approximation is valid for the light clusters only. For the heavy clusters, one needs to consider their internal structure together with the neutron-to-proton ratio which eventually can affect cluster preformation in the parent nuclei and consequently the half-life \cite{Wei2017prc}. With this in view, a separate study considering internal structure is required for the clusters with the mass number $A_c$$>$28.

\section{Results and discussion}
Considering the role of the shell effect on cluster decay, we divide our region of interest (56$\leq$Z$\leq$120) into two parts: (i) the trans-tin region in which daughter nuclei are reckoned around doubly magic nucleus $^{100}$Sn, and (ii) the trans-lead and superheavy regions where daughter nuclei are counted around doubly magic nucleus $^{208}$Pb. For the trans-tin region only one data for the cluster ($^{114}$Ba) is available \cite{nature,Guglielmetti1997PRC}, therefore, to fit the formula for this region we include 199 GLDM data \cite{nature} for cluster decay. For the $\alpha$-decay fitting, we use 107 experimental data taken from NUBASE2020 \cite{audi20201}. On the other side, for the trans-lead region, we fit the formula for 37 parent-cluster combinations (total of 61 data points due to multiple $Q$-value corresponding to various detection systems) \cite{Bonetti2007,Gupta1994,Price1989,Santhosh2012,Hourani1989} and 308 $\alpha$-decay data from NUBASE2020 \cite{audi20201}, separately. In this way, the same ITM formula can be utilized to estimate both clusters and $\alpha$-decay half-lives for both regions (trans-tin and trans-lead) which are speculated to show different decay properties of nuclei \cite{dengplb2021,saxenanew2023}. The corresponding coefficients are mentioned in Table \ref{tab:coefficient} for trans-tin region (ITM-CT and ITM-AT) and for trans-lead region (ITM-CL and ITM-AL), separately. For the nomenclature of the formula 'C' and 'A' refer to cluster and alpha decay whereas last letters 'T' and 'L' refer to trans-tin and trans-lead region, respectively (please see Table \ref{tab:coefficient}).
Since, in heavy and superheavy regions, uncertainties play a crucial role therefore we have calculated all the fitting parameters with their respective uncertainties by taking into account experimental uncertainties \cite{audi20201}. These uncertainties in parameters are mentioned in Table \ref{tab:coefficient} for all the present formulas i.e. ITM-CT, ITM-AT, ITM-CL, and ITM-AL which lead to uncertainties in the theoretical half-live with $\pm0.71$, $\pm0.40$, $\pm0.56$, and $\pm0.41$, respectively. \par
   \begin{table}[!htbp]
 \caption{The RMSE and ${\chi}^2$ of various formulas viz. ITM (present work), IUF \cite{UF2022}, ISEF \cite{cheng2022}, MBKAG \cite{jain2023}, UDL \cite{UDL2009}, TM \cite{Tavares2013},  and New UDL \cite{soylu2021} for 62 cluster decay data.}
 \centering \def\arraystretch{0.4} \resizebox{0.5\textwidth}{!}{ {\begin{tabular}{l|ccc}
 \hline
 \multicolumn{1}{c|}{Formula}&
 \multicolumn{1}{c}{No. of}&
 \multicolumn{1}{c}{RMSE}&
 \multicolumn{1}{c}{${\chi}^2$}\\
 \multicolumn{1}{c|}{}&
 \multicolumn{1}{c}{coefficient}&
 \multicolumn{1}{c}{}&
 \multicolumn{1}{c}{}\\
 \hline
 ITM (present work)      & 6+6   & 0.66    & 0.55  \\
 IUF \cite{UF2022}       & 7   & 0.87    & 0.85  \\
 ISEF \cite{cheng2022}   & 4   & 0.99    & 1.04  \\
 MBKAG \cite{jain2023}   & 5   & 1.14    & 1.41  \\
 UDL \cite{UDL2009}      & 3   & 1.52    & 2.44  \\
 TM \cite{Tavares2013}   & 4   & 1.81    & 3.51  \\
 New UDL \cite{soylu2021}& 6   & 1.94    & 4.16  \\
 \hline \end{tabular}}} \label{tab:rmse} \end{table}

\begin{figure*}[!htbp]
\centering
\includegraphics[width=1.00\textwidth]{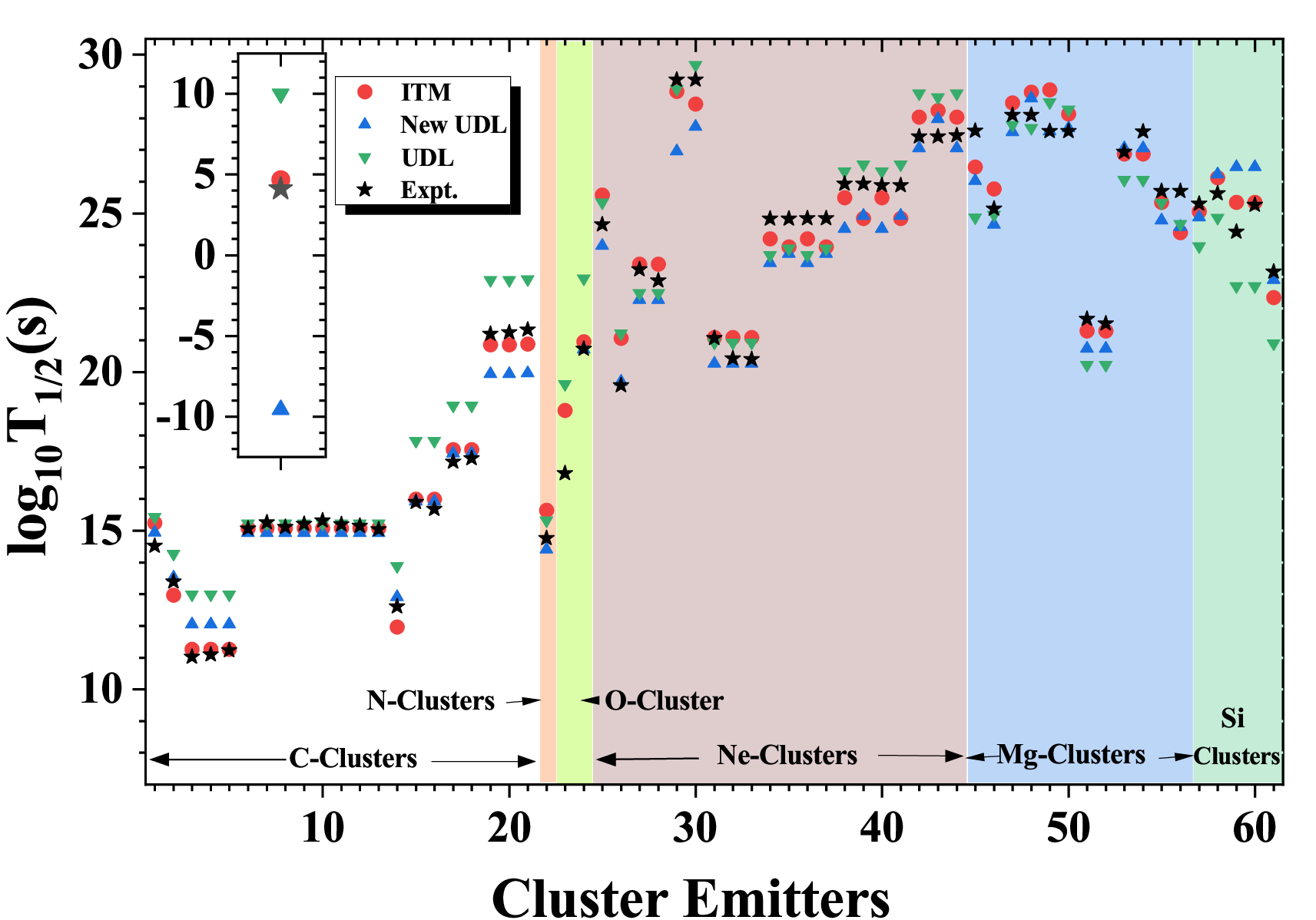}
\caption{(Colour online) Comparison of logarithmic half-lives of cluster decay, calculated by using ITM (present work), New UDL \cite{soylu2021}, and UDL \cite{UDL2009} formulas with available experimental data. X-axis refers to the clusters for which experimental data are available. Inset shows one data for trans-tin region i.e. decay of $^{12}$C cluster from $^{114}$Ba.}
\label{comparison}
\end{figure*}

\begin{figure*}[!htbp]
\centering
\includegraphics[width=1.00\textwidth]{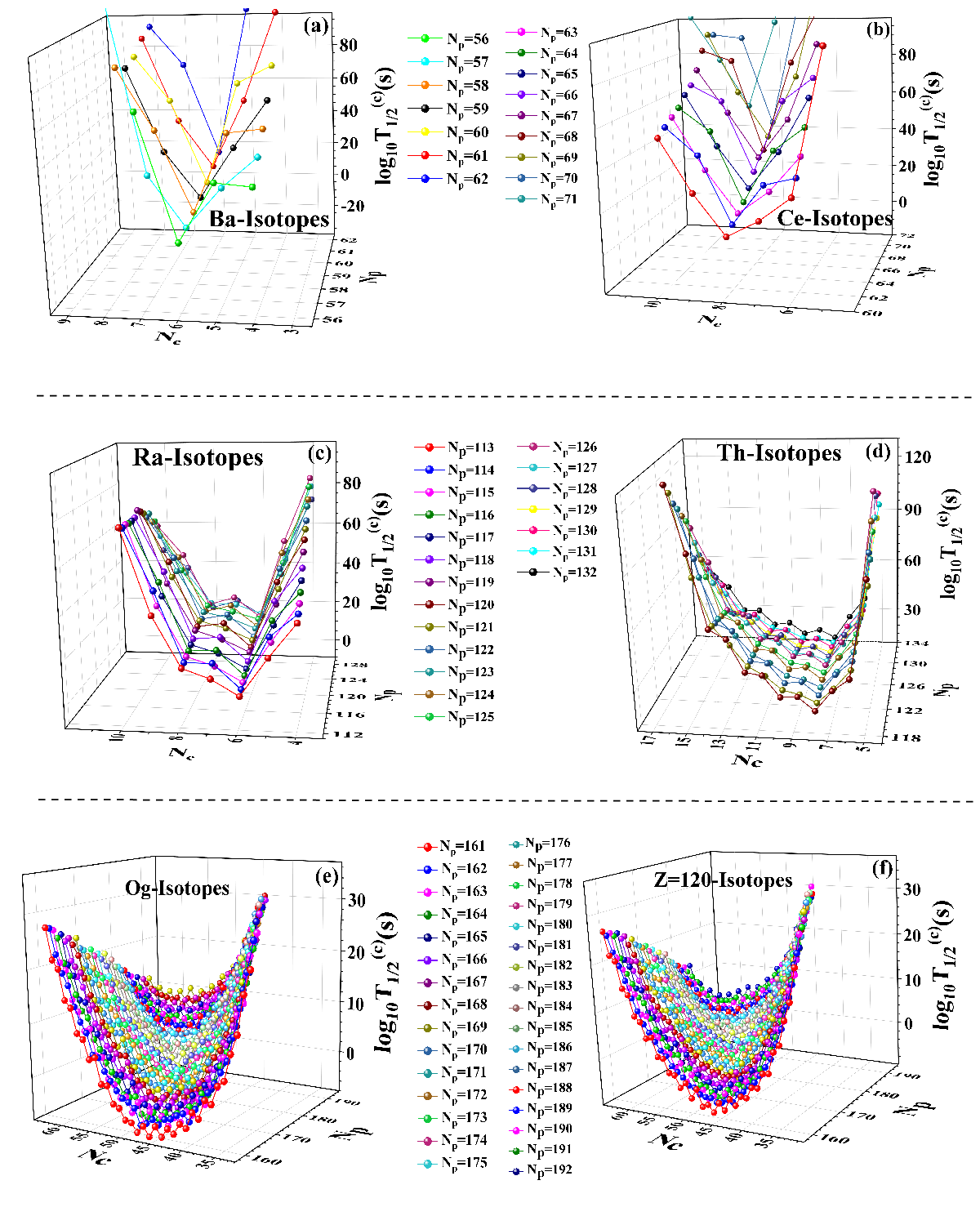}
\caption{(Colour online) Cluster decay half-lives for various isotopes of Ba (N$_{p}$=56-62), Ce (N$_{p}$=61-71), Ra (N$_{p}$=113-128), Th (N$_{p}$=120-132), Og (N$_{p}$=162-192), and Z=120 (N$_{p}$=166-192) calculated by the ITM formula.}
\label{half-lives}
\end{figure*}

The applicability of this kind of separate fitting for trans-tin and trans-lead regions can be judged by comparing the results of ITM formula with the few latest and well-established formulas of cluster decay. In Table \ref{tab:rmse}, we list the calculated values of root mean square error (RMSE) as well as $\chi^2$ (formulas are mentioned in Ref. \cite{jain2023})for ITM and other similar formulas viz. improved unified formula (IUF) \cite{UF2022}, improved semi-empirical formula (ISEF) \cite{cheng2022}, MBKAG \cite{jain2023}, UDL \cite{UDL2009}, TM  formula\cite{Tavares2013}, and New UDL \cite{soylu2021} formulas. The lowest value of RMSE of ITM formula among all the considered formulas clearly demonstrates its accuracy and improvement over the original TM formula (as RMSE improves from 1.81 to 0.61). In addition to this, the least value of $\chi^2$ exhibits pertinency of ITM formula on the ground of the number of parameters. Further, to analyze the estimation of half-lives using the ITM formula with the available experimental data, we have plotted half-lives of several clusters calculated by using the ITM formula, New UDL \cite{soylu2021}, and UDL formula \cite{UDL2009} along with their experimental half-lives \cite{Bonetti2007,Gupta1994,Price1989,Santhosh2012,Hourani1989} in Fig. \ref{comparison}. The figure contains all the experimental C, N, O, Ne, Mg, and Si clusters (in different colour bands) that belong to the trans-lead region. In the inset, we have depicted the half-life of $^{12}$C cluster from the only trans-tin parent nucleus $^{114}$Ba. A reasonable match of theoretical half-lives from ITM formula with the experimental half-lives endorses the utilization of ITM formula for the unknown region to find the possibility of cluster decay for a wide range of periodic chart from trans-tin to trans-lead as well as the superheavy regions. For C clusters in Fig. \ref{comparison}, the theoretical half-lives of all formulas are found the same for all the points due to the same $Q$-value (31.85 MeV) for the single parent-cluster combination $^{223}Ra\longrightarrow^{209}Pb+^{14}C$.\par

To explore the full periodic chart, we perform a systematic and comprehensive analysis for all the parent isotopes in the range 56$\leq$Z$\leq$120 using ITM formula by varying their corresponding daughters in such a way that cluster decay leads daughter nuclei around $^{100}$Sn for the trans-tin region and around $^{208}$Pb for rest of the chart (including trans-lead and superheavy regions). As per Eqn. (\ref{eqmtm}), half-lives of cluster decay mainly depend on $Q$-value which is calculated by using the following relation:
\begin{eqnarray}
 Q (MeV) &=& B.E.(d) + B.E.(c) - B.E.(p) \nonumber\\
&& + k[Z_{p}^{\beta}- Z_{d}^{\beta}],
 \label{Q}
 \end{eqnarray}
here, the term $k[Z_{p}^{\epsilon}-Z_{d}^{\epsilon}]$ indicates the screening effect caused by the surrounding electrons around the nucleus \cite{Denisov2009prc} where k=8.7 eV and $\epsilon$=2.517 for Z $\geq$ 60 and k=13.6 eV and $\epsilon$ =2.408 for Z $<$ 60 have been deducted from the data shown by Huang \textit{et al.} \cite{huang1976}. For accurate prediction of theoretical $Q$-values, we have used binding energies (for daughter(d), cluster(c), and parent(p) nuclei) from Weizsacker-Skyrme mass model (WS4) \cite{ws42014} which was found more precise \cite{jain2023} compared to other theories viz. Finite Range Droplet Model (FRDM) \cite{moller2019}, nonrelativistic Skyrme Hartree-Fock-Bogoliubov (HFB) \cite{hfb2004}, and relativistic mean-field theory (RMF) \cite{saxenaPLB2019,saxena2017,singh2020,saxenaplb2017}.\par
\begin{table*}[!htbp]
\caption{The comparison of branching ratio (BR) with available experimental results \cite{audi20201,Bonetti2007,Gupta1994,soylu2021} for various types of decay viz. cluster decay, $\alpha$-decay, and SF.}
\centering
\def\arraystretch{1.0}
\resizebox{1.0\textwidth}{!}{%
{\begin{tabular}{ccccScccSS}
\hline
\multicolumn{1}{c}{Parent}&
 \multicolumn{1}{c}{Emitted}&
 \multicolumn{1}{c}{Daughter}&
 \multicolumn{3}{c}{Expt. BR}&
 \multicolumn{1}{c}{}&
 \multicolumn{3}{c}{Theoretical BR}\\
 \cline{4-6}\cline{8-10}

 \multicolumn{1}{c}{nucleus}&
 \multicolumn{1}{c}{cluster}&
 \multicolumn{1}{c}{nucleus}&
 \multicolumn{1}{c}{Cluster}&
 \multicolumn{1}{c}{$\alpha$}&
 \multicolumn{1}{c}{SF}&
 \multicolumn{1}{c}{}&
 \multicolumn{1}{c}{Cluster}&
 \multicolumn{1}{c}{$\alpha$}&
 \multicolumn{1}{c}{SF}\\

\hline

\hline
 $^{221}$Fr&$^{14}$C&$^{207}$Tl&8.8e-11                 &100.00 &0.0048      &&8e-15    &100.00   &0.00\\
 $^{221}$Ra&$^{14}$C&$^{207}$Pb&1.2e-10                 &100.00 &-           &&9e-09    &100.00   &0.00\\
 $^{222}$Ra&$^{14}$C&$^{208}$Pb&3.0e-8                  &100.00 &-           &&9e-09    &100.00   &0.00\\
 $^{223}$Ra&$^{14}$C&$^{209}$Pb&8.9e-8                  &100.00 &-           &&4e-08    &100.00   &0.00\\
 $^{223}$Ac&$^{14}$C&$^{209}$Bi&-                       &99.00  &-           &&4e-07    &100.00   &0.00\\
 $^{223}$Ac&$^{15}$N&$^{208}$Pb&-                       &99.00  &-           &&8e-11    &100.00   &0.00\\
 $^{224}$Ra&$^{14}$C&$^{210}$Pb&4.0e-9                  &100.00 &-           &&6e-07    &100.00   &0.00\\
 $^{225}$Ac&$^{14}$C&$^{211}$Bi&5.3e-10                 &100.00 &-           &&5e-09    &100.00   &0.00\\
 $^{226}$Ra&$^{14}$C&$^{212}$Pb&2.6e-9                  &100.00 &-           &&2e-05    &100.00   &0.00\\
 $^{226}$Th&$^{18}$O&$^{208}$Pb&$<$3.2e-12              &100.00 &-           &&3e-11    &100.00   &0.00\\
 $^{228}$Th&$^{20}$O&$^{208}$Pb&1.13e-11                &100.00 &-           &&3e-08    &100.00   &0.00\\
 $^{230}$Th&$^{24}$Ne&$^{206}$Hg&5.8e-11                &100.00 &4e-12       &&4e-10    &100.00   &0.00\\
 $^{230}$U &$^{22}$Ne&$^{208}$Pb&4.8e-12                &100.00 &$?$         &&6.18e-13 &100.00   &0.00\\
 $^{231}$Pa&$^{24}$Ne&$^{207}$Tl&13.4e-10               &100.00 &$<$3e-10    &&6.18e-13 &100.00   &0.00\\
 $^{231}$Pa&$^{24}$Ne&$^{207}$Tl&13.4e-10               &100.00 &$<$3e-10    &&4.91e-11 &100.00   &0.00\\
 $^{232}$Th&$^{24}$Ne&$^{208}$Hg&$<$2.78e-10            &100.00 &1.1e-9      &&3.02e-09 &100.00   &0.00\\
 $^{232}$Th&$^{26}$Ne&$^{206}$Hg&$<$2.78e-10            &100.00 &1.1e-9      &&3.02e-09 &100.00   &0.00\\
 $^{232}$U &$^{24}$Ne&$^{208}$Pb&8.9e-10                &100.00 &2.7e-12     &&6.18e-10 &100.00   &0.00\\
 $^{233}$U &$^{24}$Ne&$^{209}$Pb&7.2e-11                &100.00 &$<$6e-11    &&5.21e-10 &100.00   &0.00\\
 $^{233}$U &$^{25}$Ne&$^{208}$Pb&7.2e-13                &100.00 &$<$6e-11    &&5.21e-10 &100.00   &0.00\\
 $^{233}$U &$^{28}$Mg&$^{205}$Hg&$<$1.3e-13             &100.00 &$<$6e-11    &&2.85e-12 &100.00   &0.00\\
 $^{234}$U &$^{24}$Ne&$^{210}$Pb&9e-12                  &100.00 &1.64e-9     &&5.85e-11 &100.00   &0.00\\
 $^{234}$U &$^{26}$Ne&$^{208}$Pb&9e-12                  &100.00 &1.64e-9     &&2.69e-10 &99.97    &0.03\\
 $^{234}$U &$^{28}$Mg&$^{206}$Hg&1.4e-11                &100.00 &1.64e-9     &&3.07e-11 &99.97    &0.03\\
 $^{235}$U &$^{24}$Ne&$^{211}$Pb&8e-10                  &100.00 &7e-9        &&2.88e-12 &100.00   &0.00\\
 $^{235}$U &$^{25}$Ne&$^{210}$Pb&8e-10                  &100.00 &7e-9        &&1.78e-12 &100.00   &0.00\\
 $^{235}$U &$^{28}$Mg&$^{207}$Hg&8e-10                  &100.00 &7e-9        &&1.01e-12 &100.00   &0.00\\
 $^{235}$U &$^{29}$Mg&$^{206}$Hg&8e-10                  &100.00 &7e-9        &&4.61e-13 &100.00   &0.00\\
 $^{235}$U &$^{24}$Ne&$^{211}$Pb&8e-10                  &100.00 &7e-9        &&2.87e-12 &100.00   &0.00\\
 $^{236}$U &$^{28}$Mg&$^{208}$Hg&2e-13                      &100.00 &9.4e-8      &&2.23e-12 &99.58    &0.42\\
 $^{236}$U &$^{30}$Mg&$^{206}$Hg&2e-13                      &100.00 &9.4e-8      &&1.31e-11 &99.58    &0.42\\
 $^{236}$Pu&$^{28}$Mg&$^{208}$Pb&2e-12                  &100.00 &1.9e-7      &&6.69e-12 &100.00   &0.00\\
 $^{237}$Np&$^{30}$Mg&$^{207}$Tl&$<$4e-12               &100.00 &$<$2e-10    &&1.28e-12 &100.00   &0.00\\
 $^{238}$Pu&$^{28}$Mg&$^{210}$Pb&$\approx$6e-15         &100.00 &1.9e-7      &&1.66e-14 &99.87    &0.13\\
 $^{238}$Pu&$^{30}$Mg&$^{208}$Pb&$\approx$6e-15         &100.00 &1.9e-7      &&1.50e-13 &99.87    &0.13\\
 $^{238}$Pu&$^{32}$Si&$^{206}$Hg&$\approx$1.4e-14       &100.00 &1.9e-7      &&3.28e-14 &99.87    &0.13\\
 $^{240}$Pu&$^{34}$Si&$^{206}$Hg&$<$1.3e-11             &100.00 &5.796e-6    &&2.25e-13 &93.28    &6.72\\
 $^{241}$Am&$^{34}$Si&$^{207}$Tl&$<$2.6e-13             &100.00 &3.6e-10     &&2.08e-14 &100.00   &0.00\\
 $^{242}$Cm&$^{34}$Si&$^{208}$Pb&1.1e-14                &100.00 &6.2e-6      &&5.54e-14 &99.75    &0.25\\
 \hline

\end{tabular}}
}
\label{tab:competition-decay-modes}
\end{table*}

As representative examples for the trans-tin region, in Fig. \ref{half-lives} (a) and (b), we have shown the estimated half-lives of Ba and Ce isotopes by choosing experimentally known isotopes of C and O as clusters, where each line represents half-lives of all considered clusters from the corresponding one parent isotope. Similarly, for the trans-lead region, we have shown plots (c) and (d) for Ra and Th isotopes for the emission of isotopes of C and O, respectively. Towards the superheavy region, heavy clusters Kr and Sr isotopes are estimated in plots (e) and (f) for Z=118 (Og) and Z=120 isotopes, respectively. A parabolic trend is clearly visible from all the plots and with all the extensive analysis in the range 56$\leq$Z$\leq$120 (not shown here), from which one can select a minimum (indicating least half-life) of each parabola corresponding to the potential cluster nucleus for a particular parent nucleus. From this selection, we have extracted the most probable cluster for each considered isotope in the chosen range. \par
\begin{table*}[!htbp]
\caption{Competition among various decay modes viz. cluster decay, $\alpha$-decay, and $\beta$-decay for trans-tin nuclei in the range 56$\leq$Z$\leq$81. Experimental BR and half-life are taken from NUBASE2020 \cite{audi20201} and \cite{Guglielmetti1997PRC}, respectively. Cluster decay, $\alpha$-decay half-lives are calculated by using ITM-CT and ITM-AT formulas, respectively (present work), and for the $\beta$-decay half-lives formula given by Fiset and Nix \cite{Fiset1972} has been used.}
\centering
\def\arraystretch{1.0}
\resizebox{1.1\textwidth}{!}{%
{\begin{tabular}{c|cl|ccccSSS|SSS}
\hline
\multicolumn{1}{c|}{Parent}&
\multicolumn{2}{c|}{Expt. Data}&
\multicolumn{10}{c}{Theoretical Estimates for Cluster Emission} \\
\cline{2-3}\cline{4-13}
\multicolumn{1}{c|}{Nucleus}&
\multicolumn{1}{c}{logT$_{1/2}$ }&
\multicolumn{1}{c|}{BR}&
\multicolumn{1}{c}{Emitted}&
\multicolumn{1}{c}{Daughter}&
\multicolumn{1}{c}{$Q$-value}&
\multicolumn{1}{c}{$l$}&
\multicolumn{3}{c|}{logT$_{1/2}$ (s)}&
\multicolumn{3}{c}{BR (in $\%$)}\\
\cline{8-10}\cline{11-13}
\multicolumn{1}{c|}{}&
\multicolumn{1}{c}{(s)}&
\multicolumn{1}{c|}{(in $\%$)}&
\multicolumn{1}{c}{Cluster}&
\multicolumn{1}{c}{Nucleus}&
\multicolumn{1}{c}{(MeV)}&
\multicolumn{1}{c}{}&
\multicolumn{1}{c}{Cluster}&
\multicolumn{1}{c}{$\alpha$}&
\multicolumn{1}{c|}{$\beta$}&
\multicolumn{1}{c}{Cluster}&
\multicolumn{1}{c}{$\alpha$}&
\multicolumn{1}{c}{$\beta$}\\

\hline
$^{114}$Ba&$>$4.10&$\beta^{+}\approx$100;$\beta^{+}$p=20;&$^{12}$C&$^{102}$Sn&19.00&0&4.68& 2.48 & -0.34 & 0.0010 & 0.1431 & 99.8560\\
&&$\alpha$=0.9;$^{12}$C$<$0.0034&&&  & &  &  &&  &  & \\
$^{111}$La&-&-&$^{13}$N&$^{98}$Sn  & 25.64 & 1 & 1.27  & -5.00 & 2.19 & 0.0001 & 99.9999 & 0.0000\\
$^{112}$La&-&-&$^{13}$N&$^{99}$Sn  & 26.15 & 3 & 0.26  & -3.09 & 1.84 & 0.0438 & 99.9550 & 0.0012\\
$^{113}$La&-&-&$^{13}$N&$^{100}$Sn & 26.61 & 1 & -0.51 & -1.33 & 2.41 & 13.1268& 86.8573 & 0.0160\\

$^{112}$Ce&-&-       & $^{17}$O & $^{95}$Sn & 32.55 & 2 & 0.85  & -5.12 & 2.77 & 0.0001  & 99.9999 & 0.0000\\
$^{113}$Ce&-&-       & $^{17}$O & $^{96}$Sn & 33.58 & 2 & -1.05 & -3.41 & 2.13 & 0.4308  & 99.5689 & 0.0003\\
$^{114}$Ce&-&-       & $^{17}$O & $^{97}$Sn & 34.73 & 2 & -3.04 & -2.79 & 2.95 & 63.9793 & 36.0206 & 0.0001\\
$^{115}$Ce&-&-       & $^{17}$O & $^{98}$Sn & 35.88 & 2 & -4.84 & -2.76 & 2.36 & 99.1812 & 0.8188  & 0.0000\\
$^{116}$Ce&-&-       & $^{16}$O & $^{100}$Sn& 32.30 & 0 & 1.92  & -2.65 & 3.29 & 0.0027  & 99.9972 & 0.0001\\
$^{117}$Ce&-&-       & $^{16}$O & $^{101}$Sn& 31.45 & 3 & 5.06  & -0.52 & 2.58 & 0.0003  & 99.9208 & 0.0790\\
$^{118}$Ce&-&-       & $^{16}$O & $^{102}$Sn& 30.07 & 0 & 5.71  &  3.22 & 3.48 & 0.2104  & 64.2798 & 35.5098\\

$^{114}$Pr &-&-&  $^{15}$F& $^{99}$Sn  & 37.38 & 4 &  2.62 & -2.90 & 1.63 &0.0003  & 99.9967 &0.0030\\
$^{115}$Pr &-&-&  $^{15}$F& $^{100}$Sn & 37.04 & 1 &  4.50 & -2.44 & 2.11 &0.0001  & 99.9953 &0.0046\\
$^{116}$Pr &-&-&  $^{17}$F& $^{99}$Sn  & 36.65 & 6 &  4.74 & -1.51 & 1.75 &0.0001  & 99.9460 &0.0540\\
$^{117}$Pr &-&-&  $^{17}$F& $^{100}$Sn & 36.92 & 1 &  5.53 & -1.10 & 2.33 &0.0003  & 99.9383 &0.0614\\

$^{116}$Nd &-&- & $^{18}$Ne& $^{98}$Sn  & 43.32 & 0 &  3.63 & -0.54 & 2.70 &0.0067  & 99.9357 &0.0576\\
$^{117}$Nd &-&- & $^{18}$Ne& $^{99}$Sn  & 43.11 & 2 &  5.87 & 1.22  & 2.08 &0.0020  & 87.8970 &12.1011\\
$^{118}$Nd &-&- & $^{18}$Ne& $^{100}$Sn & 42.68 & 0 &  5.42 & 2.30  & 2.86 &0.0584  & 78.4906 &21.4510\\
$^{121}$Sm & -& -  &   $^{22}$Mg& $^{99}$Sn  & 54.21 & 0 &  7.51 & 4.23 & 1.94 &0.0003 &0.5094 &99.4903\\

\hline
\end{tabular}}
}
\label{tab:competition-trans-tin}
\end{table*}

This systematic study indeed provides the most probable cluster for each considered isotope, however, its lowest half-life does not guarantee the chances of cluster decay due to its contest with other probable decay modes. For example, as per experimental systematics, $\beta$-decay and $\alpha$-decay are the principal decay modes for the trans-tin region whereas $\alpha$-decay and spontaneous fission (SF) play the primary roles in the trans-lead and superheavy region. As a result, the chances of cluster decay are always dependent on its competition with other primary decay modes. With this in view, we have calculated half-lives of other decay modes using (i) ITM-AT and ITM-AL formulas (mentioned in Table \ref{tab:coefficient}) for the estimation of half-lives of $\alpha$-decay in trans-tin and trans-lead region, respectively, (ii) Fisset and Nix formula \cite{Fiset1972} for $\beta$-decay in trans-tin region, and (iii) Modified Bao Formula (MBF) \cite{Saxena2021jpg} for spontaneous fission in trans-lead and superheavy regions. These formulas for $\beta$-decay and SF are recently been applied and found very successful in determining the respective half-lives in the heavy and superheavy regions \cite{singh2020,Saxena2021jpg,jain2023PS,rsharma2022,saxena2021QF}. The competition can be quantized in the form of branching ratios (BR) which are defined as:
\begin{equation}\label{eq:branching}
  BR  = \frac{T^{Th.}_{1/2}}{T^{Cluster/\alpha/\beta/SF}_{1/2}},
\end{equation}
where, $T^{Th.}_{1/2}$ is the total half-life calculated by considering half-lives of all decay modes (cluster, $\alpha$, and $\beta$/SF), and the relation is given by:
\begin{equation}\label{eq:Total-hl}
  \frac{1}{T^{Th.}_{1/2}} = \frac{1}{T^{Cluster}_{1/2}}+\frac{1}{T^{\alpha}_{1/2}}+\frac{1}{T^{\beta/SF}_{1/2}},
\end{equation}
where the superscripts refer to the half-lives of concerned decay modes. The superscript $\beta$/SF refers to the consideration of $\beta$-decay (for trans-tin region) or SF (for trans-lead and superheavy regions). A good test of our predictions of all the used formulas can be performed using Eqn. (\ref{eq:branching}) from which we can compare values of BR of various decay modes from our theoretical half-lives with the BR available in NUBASE2020 \cite{audi20201}. Taking this into consideration, we have listed the theoretical and experimental BR (in percentage) for several known cluster decay \cite{audi20201,Bonetti2007,Gupta1994,soylu2021} in Table \ref{tab:competition-decay-modes}. The dominance of $\alpha$-decay can be easily verified by Table \ref{tab:competition-decay-modes} because the value of concern BR is near to 100$\%$. In contrast, the values of BR for cluster decay are very less, however, the experimental values of BR for cluster decay are indeed reproduced very well from the half-life obtained by using the ITM formula. This excellent match endorses our approach of determining BR by using ITM and other formulas for our considered unknown domain of the periodic chart.

In Table \ref{tab:competition-trans-tin}, a worthful competition among various decay modes viz. cluster decay, $\alpha$-decay, and $\beta$-decay is listed for the trans-tin region in the range 56$\leq$Z$\leq$81. As mentioned earlier this whole systematic study provides many clusters i.e. one probable cluster for each considered isotope (as obtained by the minimum of each parabola of Fig. \ref{half-lives}) but only those are mentioned in the Table \ref{tab:competition-trans-tin} which are found with reasonable BR as compared to other decay modes. More than 15 cluster decays are found probable in the trans-tin region, and in the future, looking into the experimental progress one may expect more observation of cluster decay in the trans-tin region where so far only one candidate i.e. $^{114}$Ba is detected. It is noticeable from Table \ref{tab:competition-trans-tin} that two mentioned clusters have values of BR more than 60$\%$ or more than $\alpha$ or $\beta$ decay modes and at the same time the values of BR for a few nuclei are several orders of magnitude larger than the BR of experimentally known clusters (please see Table \ref{tab:competition-decay-modes}). Therefore, this comparison certainly anticipates cluster decay as a prominent mode of decay in several nuclei of the trans-tin region. \par

\begin{table*}[!htbp]
\caption{Same as Table \ref{tab:competition-trans-tin}, but for trans-lead and superheavy regions in the range 88$\leq$Z$\leq$120. Cluster decay and $\alpha$-decay half-lives are calculated by using ITM-CL and ITM-AL formulas, respectively (present work). For the SF half-lives MBF formula \cite{Saxena2021jpg} has been used.}
\centering
\def\arraystretch{1.0}
\resizebox{1.12\textwidth}{!}{%
{\begin{tabular}{c|cl|ccccSSS|SSS}
\hline
\multicolumn{1}{c|}{Parent}&
\multicolumn{2}{c|}{Expt. Data}&
\multicolumn{10}{c}{Theoretical Estimates for Cluster Emission} \\
\cline{2-3}\cline{4-13}
\multicolumn{1}{c|}{Nucleus}&
\multicolumn{1}{c}{logT$_{1/2}$}&
\multicolumn{1}{c|}{BR}&
\multicolumn{1}{c}{Emitted}&
\multicolumn{1}{c}{Daughter}&
\multicolumn{1}{c}{$Q$-value}&
\multicolumn{1}{c}{$l$}&
\multicolumn{3}{c|}{logT$_{1/2}$ (s)}&
\multicolumn{3}{c}{BR (in $\%$)}\\
\cline{8-10}\cline{11-13}
\multicolumn{1}{c|}{}&
\multicolumn{1}{c}{(s)}&
\multicolumn{1}{c|}{(in $\%$)}&
\multicolumn{1}{c}{Cluster}&
\multicolumn{1}{c}{Nucleus}&
\multicolumn{1}{c}{(MeV)}&
\multicolumn{1}{c}{}&
\multicolumn{1}{c}{Cluster}&
\multicolumn{1}{c}{$\alpha$}&
\multicolumn{1}{c|}{SF}&
\multicolumn{1}{c}{Cluster}&
\multicolumn{1}{c}{$\alpha$}&
\multicolumn{1}{c}{SF}\\

\hline
$^{285}$Mc &-&- &$^{81}$As& $^{204}$Pb & 272.66 &2 &  2.95& -0.91 & 6.02 &0.01 & 99.99&0.00\\
$^{289}$Mc &-&-&$^{81}$As& $^{208}$Pb & 272.95 &2 &  4.33 &  0.16  & 7.91 &0.01 & 99.99&0.00\\
$^{275}$Lv &-&-&$^{80}$Se& $^{195}$Pb & 284.92 &6 &  -3.26& -5.81  & -5.24 &0.22 & 78.64&21.14\\
$^{276}$Lv &-&-&$^{80}$Se& $^{196}$Pb & 284.87 &0 &  -5.09& -7.47  & -6.05 &0.40 & 95.92&3.69\\
$^{277}$Lv &-&-&$^{80}$Se& $^{197}$Pb & 284.48 &2 &  -3.65& -6.19  & -1.04 &0.29 & 99.71&0.00\\
$^{278}$Lv &-&-&$^{80}$Se& $^{198}$Pb & 284.28 &0 &  -4.02& -6.77 & -5.64 &0.16 & 92.93&6.91\\
$^{279}$Lv &-&-&$^{80}$Se& $^{199}$Pb & 284.06 &6 &  -1.23& -4.93  & -1.86 &0.02 & 99.89&0.09\\
$^{280}$Lv &-&-&$^{80}$Se& $^{200}$Pb & 283.90 &0 &  -3.04& -6.71  & -2.25 &0.02 & 99.98&0.00\\
$^{281}$Lv &-&-&$^{80}$Se& $^{201}$Pb & 283.53 &1 &  -2.00& -3.64  & 1.02  &2.25 & 97.75&0.00\\
$^{282}$Lv &-&-&$^{80}$Se& $^{202}$Pb & 283.39 &0 &  -2.04& -5.68  & -0.45 &0.02 & 99.98&0.00\\
$^{283}$Lv &-&-&$^{82}$Se& $^{201}$Pb & 283.49 &1 &  -1.18& -4.30  & 3.33  &0.07 & 99.93&0.00\\
$^{284}$Lv &-&-&$^{82}$Se& $^{202}$Pb & 283.64 &0 &  -1.33& -4.63  & 1.44  &0.05 & 99.95&0.00\\
$^{285}$Lv &-&-&$^{82}$Se& $^{203}$Pb & 283.48 &1 &  -0.39& -3.19  & 4.69  &0.16 & 99.84&0.00\\
$^{286}$Lv &-&-&$^{82}$Se& $^{204}$Pb & 283.41 &0 &  -0.47& -3.55  & 2.97  &0.08 & 99.92&0.00\\
$^{287}$Lv &-&-&$^{82}$Se& $^{205}$Pb & 283.44 &1 &   0.40& -3.47  & 5.86  &0.01 & 99.99&0.00\\
$^{288}$Lv &-&-&$^{82}$Se& $^{206}$Pb & 283.60 &0 &   0.22& -3.47  & 3.40  &0.02 & 99.98&0.00\\
$^{289}$Lv &-1.80&$\alpha$$?$&$^{82}$Se& $^{207}$Pb & 283.43 &3 &   1.88& -2.35 & 6.64  &0.01 &99.99&0.00\\
$^{290}$Lv &-2.05$\pm$0.15&$\alpha$$\approx$100;SF$?$&$^{84}$Se&$^{206}$Pb&283.57&0&0.97&-3.00&4.41&0.01 & 99.99&0.00\\
$^{291}$Lv &-1.59$\pm$0.22&$\alpha$$\approx$100; SF$?$&$^{84}$Se& $^{207}$Pb & 283.77 &1 &   1.77& -2.20  & 7.59  &0.01 & 99.99&0.00\\
$^{279}$Ts &-&-                              &$^{83}$Br& $^{196}$Pb & 294.22 &0 &  -6.32& -6.37  & -0.88 &47.04 & 52.96&0.00\\
$^{281}$Ts &-&-                              &$^{83}$Br& $^{198}$Pb & 293.88 &0 &  -5.35& -6.13  & 2.40  &14.32 & 85.68&0.00\\
$^{283}$Ts &-&-                              &$^{83}$Br& $^{200}$Pb & 293.58 &0 &  -4.42& -5.54  & 4.00  &7.13  & 92.87&0.00\\
$^{285}$Ts &-&-                              &$^{83}$Br& $^{202}$Pb & 293.17 &0 &  -3.48& -4.73  & 5.98  &5.25  & 94.75&0.00\\
$^{287}$Ts &-&-                              &$^{85}$Br& $^{202}$Pb & 292.75 &0 &  -2.54& -3.97  & 7.35  &3.58  & 96.42&0.00\\
$^{289}$Ts &-&-                              &$^{85}$Br& $^{204}$Pb & 292.92 &0 &  -1.84& -3.81  & 8.50  &1.04  & 98.96&0.00\\
$^{291}$Ts &-2.70&$\alpha$$?$$;$SF$?$        &$^{85}$Br& $^{206}$Pb & 293.18 &0 &  -1.18& -3.26  & 10.10 &0.82  & 99.18&0.00\\
$^{293}$Ts &-1.60$\pm$0.11&$\alpha$=100      &$^{85}$Br& $^{208}$Pb & 292.85 &0 &  -0.33& -3.04 & 9.45  &0.20  & 99.80&0.00\\
$^{295}$Ts &-&-                              &$^{87}$Br& $^{208}$Pb & 291.24 &2 &   1.83& -2.34 & 10.13 &0.01  & 99.99&0.00\\
$^{280}$Og &-&-                              &$^{84}$Kr& $^{196}$Pb & 305.72 &0 &   -9.19& -7.67  & -1.14&97.09  & 2.91 &0.00\\
$^{281}$Og &-&-                              &$^{86}$Kr& $^{195}$Pb & 304.59 &1 &   -7.87& -5.80  & -1.02&99.16  & 0.84 &0.00\\
$^{282}$Og &-&-                              &$^{84}$Kr& $^{198}$Pb & 304.93 &0 &   -8.06& -7.29 & -1.80&85.57  & 14.43&0.00\\
$^{283}$Og &-&-                              &$^{84}$Kr& $^{199}$Pb & 304.54 &1 &   -7.01& -4.37  & 1.54 &99.77  & 0.23 &0.00\\
$^{284}$Og &-&-                              &$^{84}$Kr& $^{200}$Pb & 304.28 &0 &   -7.01& -6.82 & -0.25&60.71  & 39.29&0.00\\
$^{285}$Og &-&-                              &$^{86}$Kr& $^{199}$Pb & 304.17 &1 &   -6.06& -6.54 & 3.42 &24.99  & 75.01&0.00\\
$^{286}$Og &-&-                              &$^{86}$Kr& $^{200}$Pb & 304.14 &0 &   -6.15& -6.26 & 1.59 &43.80  & 56.20&0.00\\
$^{287}$Og &-&-                              &$^{86}$Kr& $^{201}$Pb & 303.98 &1 &   -5.20& -6.04 & 5.05 &12.50  & 87.50&0.00\\
$^{288}$Og &-&-                              &$^{84}$Kr& $^{204}$Pb & 302.62 &0 &   -4.83& -5.70 & 3.17 &11.90  & 88.10&0.00\\
$^{289}$Og &-&-                              &$^{86}$Kr& $^{203}$Pb & 303.76 &1 &   -4.34& -5.64 & 6.65 &4.75   & 95.25&0.00\\
$^{290}$Og &-&-                              &$^{86}$Kr& $^{204}$Pb & 303.69 &0 &   -4.43& -5.64 & 4.50 &5.83   & 94.17&0.00\\
$^{291}$Og &-&-                              &$^{86}$Kr& $^{205}$Pb & 303.54 &3 &   -2.78& -4.43 & 7.93 &2.16   & 97.84&0.00\\
$^{292}$Og &-&-                              &$^{86}$Kr& $^{206}$Pb & 303.53 &0 &   -3.62& -4.95  & 6.14 &4.44   & 95.56&0.00\\
$^{293}$Og &-3.00&$\alpha$$?$                &$^{87}$Kr& $^{206}$Pb & 302.51 &0 &   -2.89& -4.07  & 9.89 &6.21   & 93.79&0.00\\
$^{294}$Og&-3.15$\pm$0.20&$\alpha$$\approx$100$;$SF$?$&$^{86}$Kr& $^{208}$Pb & 302.91 &0 &   -2.66& -4.84  & 6.15 &0.67   & 99.33&0.00\\
$^{295}$Og &-0.175$\pm$0.47&$\alpha$$\approx$100       &$^{87}$Kr& $^{208}$Pb & 301.96 &0 &   -1.97& -4.26 & 8.45 &0.51   & 99.49&0.00\\
$^{296}$Og &-&-                              &$^{88}$Kr& $^{208}$Pb & 301.70 &0 &   -1.52& -3.95 & 5.65 &0.37   & 99.63&0.00\\
$^{299}$Og &-&-                              &$^{91}$Kr& $^{208}$Pb & 298.62 &0 &    0.62& -3.61  & 7.57 &0.01   & 99.99&0.00\\
$^{287}$120 &-&-&$^{88}$Sr& $^{199}$Pb & 325.93 &3 &   -10.67& -5.90 & 3.70 &100.00  & 0.00&0.00\\
$^{288}$120 &-&-&$^{88}$Sr& $^{200}$Pb & 325.54 &0 &   -11.35& -7.27  & 2.03 &99.99   & 0.01&0.00\\
$^{289}$120 &-&-&$^{89}$Sr& $^{200}$Pb & 324.18 &0 &   -10.50& -7.23  & 5.48 &99.95   & 0.05&0.00\\
$^{289}$120 &-&-&$^{90}$Sr& $^{199}$Pb & 323.31 &1 &    -9.71& -7.23  & 5.48 &99.67   & 0.33&0.00\\
$^{290}$120 &-&-&$^{88}$Sr& $^{202}$Pb & 324.70 &0 &   -10.27& -7.19  & 3.33 &99.92   & 0.08&0.00\\
$^{291}$120 &-&-&$^{88}$Sr& $^{203}$Pb & 324.10 &1 &    -9.17& -6.00& 7.39 &99.93   & 0.07&0.00\\
$^{292}$120 &-&-&$^{88}$Sr& $^{204}$Pb & 323.63 &0 &    -9.13& -6.78  & 5.39 &99.55   & 0.45&0.00\\
$^{293}$120 &-&-&$^{90}$Sr& $^{203}$Pb & 322.58 &1 &    -7.89& -5.79  & 9.56 &99.22   & 0.78&0.00\\
$^{294}$120 &-&-&$^{88}$Sr& $^{206}$Pb & 322.79 &0 &    -8.08& -6.38 & 7.69 &98.07   & 1.93 &0.00\\
$^{295}$120 &-&-&$^{89}$Sr& $^{206}$Pb & 322.26 &0 &    -7.53& -5.54 & 10.85&98.97   & 1.03 &0.00\\
$^{296}$120 &-&-&$^{90}$Sr& $^{206}$Pb & 322.29 &0 &    -7.16& -6.51  & 7.61 &81.86   & 18.14&0.00\\
$^{297}$120 &-&-&$^{89}$Sr& $^{208}$Pb & 321.13 &0 &    -6.40& -6.16  & 10.10&63.71   & 36.29&0.00\\
$^{298}$120 &-&-&$^{90}$Sr& $^{208}$Pb & 321.34 &0 &    -6.11& -5.91  & 5.25 &61.19   & 38.81&0.00\\
$^{299}$120 &-&-&$^{91}$Sr& $^{208}$Pb & 320.72 &2 &    -4.68& -6.32  & 7.91 &2.22    & 97.78&0.00\\
$^{300}$120 &-&-&$^{92}$Sr& $^{208}$Pb & 320.87 &0 &    -5.23& -6.40  & 4.72 &6.34    & 93.66&0.00\\
$^{301}$120 &-&-&$^{93}$Sr& $^{208}$Pb & 320.00 &0 &    -4.59& -5.10 & 7.08 &23.54   & 76.46&0.00\\
$^{302}$120 &-&-&$^{94}$Sr& $^{208}$Pb & 319.86 &0 &    -4.19& -5.65  & 3.41 &3.38    & 96.62&0.00\\
$^{303}$120 &-&-&$^{95}$Sr& $^{208}$Pb & 318.35 &1 &    -2.84& -4.63  & 5.39 &1.58    & 98.42&0.00\\
$^{304}$120 &-&-&$^{96}$Sr& $^{208}$Pb & 317.35 &0 &    -2.65& -5.39 & 1.20 &0.18    & 99.82&0.00\\
$^{305}$120 &-&-&$^{94}$Sr& $^{211}$Pb & 315.96 &0 &    -1.84& -5.39  & 0.60 &0.03    & 99.97&0.00\\

\hline
\end{tabular}}
}
\label{tab:competition-trans-lead}
\end{table*}

In a similar way, the study has been performed for trans-lead nuclei together with superheavy nuclei, and the probable clusters are mentioned in Table \ref{tab:competition-trans-lead}. Various heavy clusters are found with comparable BR in this region of the periodic chart and show a great probability of decay compared to $\alpha$-decay and SF, especially for Z$=$118 and 120 isotopes similar to what was found in Refs. \cite{zhang2018,soylu2019,santhosh2018,poenaru2018,poenaru2018EPJA,matheson2018,Warda2018,jain2023PS}. In both the tables (Tables \ref{tab:competition-trans-tin} and \ref{tab:competition-trans-lead}), the theoretical half-lives of various decay modes are mentioned in columns 8, 9, and 10 which by using Eqn. (\ref{eq:Total-hl}) provide total half-lives with an excellent agreement with the available experimental half-lives (mentioned in column 2). Hence, the prediction of various cluster decay in the present article is reliable and of great importance due to its sizeable probability over other prominent decay modes of heavy and superheavy regions.

\section{Conclusions}
An extensive analysis of cluster decay within the range 56$\leq$Z$\leq$120 is carried out using an improved version of a semi-empirical formula in which isospin and angular momentum effects are included. This improved Tavares and Medeiros formula (ITM) is found with greater accuracy when compared with the available data of cluster decay in the trans-lead region. The applicability of ITM formula is demonstrated by considering daughter nuclei around the double magic nucleus $^{100}$Sn for trans-tin nuclei and $^{208}$Pb for trans-lead and superheavy nuclei. The formula is indeed found suitable for the trans-tin and trans-lead regions including the superheavy domain of the periodic chart. Competition of cluster decay mode with other probable decay modes is visualized in the form of branching ratios (BR) which are also found in an excellent match with the available experimental data. Various clusters with substantial BR are reported separately for trans-tin, trans-lead, and superheavy regions. This kind of comprehensive analysis is expected to stimulate the wide region of the periodic chart (56$\leq$Z$\leq$120) in view of cluster decay and to enforce theoretical and experimental studies eyeing heavy-ion or cluster decays.
\section{Acknowledgement}
GS acknowledges the support provided by SERB (DST), Govt. of India under SIR/2022/000566, and would like to thank Prof. Nils Paar for his kind hospitality at the University of Zagreb, Croatia. AJ is indebted to Prof. S. K. Jain, Manipal University, Jaipur, India for his guidance.


\begin{thebibliography}{}
\bibitem{sandulescu1980} A. Sandulescu, D. Poenaru, W. Greiner, Sov. J. Part. Nucl. II. \textbf{11}, 528 (1980).
\bibitem{rose1984}  H.J. Rose, G.A. Jones, Nature \textbf{307}, 245 (1984).
\bibitem{Bonetti2007} R. Bonetti, A. Guglielmetti, Rom. Rep. Phys. \textbf{59}, 301 (2007).
\bibitem{Kumar2003} S. Kumar \textit{et al.}, J. Phys. G: Nucl. Part. Phys. \textbf{29}, 625 (2003).
\bibitem{Kumar2009} S. Kumar, R. Rani, R. Kumar, J. Phys. G: Nucl. Part. Phys. \textbf{36}, 015110 (2009).
\bibitem{Gupta2003} R.K. Gupta, \textit{et al.}, Phys. Rev. C \textbf{68}, 034321 (2003).
\bibitem{Gupta1994} R.K. Gupta, W. Greiner, Int. J. Mod. Phys. E \textbf{03}, 335 (1994).
\bibitem{Price1989} P.B. Price, Annu. Rev. Nucl. Part. Sci. \textbf{39}, 19 (1989).
\bibitem{Santhosh2012} K.P. Santhosh, B. Priyanka, M.S. Unnikrishnan, Nucl. Phys. A \textbf{889}, 29 (2012).
\bibitem{Hourani1989} E. Hourani, M. Hussonnois, D.N. Poenaru, Ann. Phys. (Paris) \textbf{14}, 311 (1989).
\bibitem{Royer2001} G. Royer, R. Moustabchir, Nucl. Phys. A \textbf{683}, 182 (2001).
\bibitem{Singh2022} A. Singh \textit{et al.}, J. Phys. G: Nucl. Part. Phys. \textbf{49}, 025101 (2022).
\bibitem{kulkin2005} S.N. Kuklin, G.G. Adamian, N.V. Antonenko, Phys. Rev. C \textbf{71}, 014301 (2005).
\bibitem{Og1994} Y.T. Oganessian \textit{et al.}, Z. Fur Phys. A Hadron. Nucl. \textbf{349}, 341 (1994).
\bibitem{Guglielmetti1995} A. Guglielmetti \textit{et al.}, Phys. Rev. C. \textbf{52}, 740 (1995).
\bibitem{Guglielmetti1995npa} A. Guglielmetti \textit{et al.},  Nucl. Phys. A \textbf{583}, 867 (1995).
\bibitem{prathapan2021} K. Prathapan, R.K. Biju, Int. J. Mod. Phys. E \textbf{30}, 2150106 (2021).
\bibitem{Santhosh2022prc} K.P. Santhosh, T.A. Jose, N. K. Deepak, Phys. Rev. C \textbf{105}, 054605 (2022).
\bibitem{Royer2022} G. Royer, Q. Ferrier, M. Pineau, Nucl. Phys. A \textbf{1021}, 122427 (2022).
\bibitem{Nithya2022} C. Nithya, K.P. Santhosh, Nucl. Phys. A \textbf{1020}, 122400 (2022).

\bibitem{poenaru2018} D.N. Poenaru, R.A. Gherghescu, Phys. Rev. C \textbf{97}, 044621 (2018).
\bibitem{poenaru2018EPJA} D.N. Poenaru, H. St{\"o}cker, R.A. Gherghescu, Eur. Phys. J. A \textbf{54}, 14 (2018).
\bibitem{matheson2018} Z. Matheson, S.A. Giuliani, W. Nazarewicz, J. Sadhukhan, N. Schunck, Phys. Rev. C \textbf{99}, 041304 (2019).
\bibitem{Warda2018} M. Warda, A. Zdeb, L.M. Robledo, Phys. Rev. C \textbf{98}, 041602(R) (2018).
\bibitem{poenaru2012} D.N. Poenaru, R.A. Gherghescu, W. Greiner, Phys. Rev. C \textbf{85}, 034615 (2012).
\bibitem{zhang2018} Y.L. Zhang, Y.Z. Wang, Phys. Rev. C  \textbf{97}, 014318 (2018).
\bibitem{santhosh2008} K.P. Santhosh, R.K. Biju, J. Phys. G: Nucl. Part. Phys. \textbf{36}, 015107 (2009).
\bibitem{santhosh2018} K.P. Santhosh, C. Nithya, Phys. Rev. C \textbf{97}, 064616 (2018).
\bibitem{soylu2019} A. Soylu, F. Koyuncu, Eur. Phys. J. A \textbf{55}, 118 (2019).
\bibitem{jain2023PS} A. Jain, P.K. Sharma, S.K. Jain, Dashty T. Akrawy, G. Saxena, Phys. Scr. \textbf{98}, 085304 (2023).
\bibitem{Poenaru1982} D.N. Poenaru, M. Ivascu, D. Mazilu, Comput. Phys. Commun. \textbf{25}, 297 (1982).
\bibitem{Delion2001} D.S. Delion, J. Suhonen, Phys. Rev. C. \textbf{64}, 064302 (2001).
\bibitem{Poenaru1996} D.N. Poenaru, W. Greiner, Clarendon Press, Oxford, (1996).
\bibitem{Poenaru1996IPPB} D.N. Poenaru, Institute of Physics Publishing, Bristol, (1996).
\bibitem{arun2009} S.K. Arun \textit{et al.}, Phys. Rev. C  \textbf{79}, 064616 (2009).
\bibitem{Brack1972} M. Brack \textit{et al.}, Rev. Mod. Phys. \textbf{44}, 320 (1972).
\bibitem{Wahl1988} A.C. Wahl, At. Data Nucl. Data Tables \textbf{39}, 1 (1988).
\bibitem{Poenaru1985}  D.N. Poenaru \textit{et al.}, Phys. Rev. C \textbf{32}, 572 (1985).
\bibitem{Poenaru1986}  D.N. Poenaru \textit{et al.}, At. Data Nucl. Data Tables \textbf{34}, 423 (1986).
\bibitem{Joshua2022} Joshua T. Majekodunmi \textit{et al.}, Phys. Rev. C \textbf{105}, 044617 (2022).
\bibitem{Yahya2022} W.A Yahya, T.T. Ibrahim, Eur. Phys. J. A \textbf{58}, 48 (2022).
\bibitem{Hosseini2023} S.S. Hosseini, S.M. Motevalli, Phys. Rev. C \textbf{107}, 034611 (2023).
\bibitem{Dagtas2022} R. Dagtas, O. Bayrak, Phys. Scr. \textbf{97}, 105301 (2022).
\bibitem{Liu2021} Hong-Ming Liu \textit{et al.}, Phys. Scr. \textbf{96}, 125322 (2021).
\bibitem{Horoi2004} Mihai Horoi, J. Phys. G: Nucl. Part. Phys. \textbf{30}, 945 (2004).
\bibitem{Balasubramaniam2004} M. Balasubramaniam \textit{et al.}, Phys. Rev. C \textbf{70}, 017301 (2004).
\bibitem{Ren2004} Z. Ren, C. Xu, Z. Wang, Phys. Rev. C \textbf{70}, 034304 (2004).
 \bibitem{NRDX2008} D. Ni, Z. Ren, T. Dong, C. Xu, Phys. Rev. C \textbf{78}, 044310 (2008).
\bibitem{UDL2009}  C. Qi, F.R. Xu, R.J. Liotta, R. Wyss, Phy. Rev. Lett. \textbf{103}, 072501 (2009).
\bibitem{Poenaru2011} D.N. Poenaru, R.A. Gherghescu, W. Greiner, Phys. Rev. C \textbf{83}, 014601 (2011).
\bibitem{Tavares2013} O.A.P. Tavares, E.L. Medeiros, Eur. Phys. J. A \textbf{49}, 6 (2013).
\bibitem{jain2023} A. Jain, P.K. Sharma, S.K. Jain, J.K. Deegwal, G. Saxena, Nucl. Phys. A \textbf{1031}, 122597 (2023).
\bibitem{soylu2021} A. Soylu, C. Qi, Nucl. Phys. A \textbf{1013}, 122221 (2021).
\bibitem{UF2022} M. Ismail, A.Y. Ellithi, A. Adela, M.A. Abbas, Eur. Phys. J. A \textbf{58}, 225 (2022).
\bibitem{Wang2021CPC} Y. Wang \textit{et al.}, Chin. Phys. C \textbf{45}, 044111 (2021).
\bibitem{cheng2022} S. Cheng \textit{et al.}, Eur. Phys. J. A \textbf{58}, 168 (2022).
\bibitem{Lin2023} Lin-Jing Qi \textit{et al.}, Chin. Phys. C  \textbf{47}, 064107 (2023).
\bibitem{nature} Y. Gao, J. Cui, Y. Wang, J. Gu, Sci Rep \textbf{10}, 9119 (2020).
\bibitem{Guglielmetti1997PRC} A. Guglielmetti \textit{et al.}, Phys. Rev. C.  \textbf{56}, R2912 (1997).
\bibitem{audi20201} F.G. Kondev \textit{et al.}, Chin. Phys. C \textbf{45}, 030001 (2021).
\bibitem{sharma2021npa} P.K. Sharma, A. Jain, G. Saxena, Nucl. Phys. A \textbf{1016}, 122318 (2021).
\bibitem{saxenajpg2023} G. Saxena \textit{et al.}, J. Phys. G: Nucl. Part. Phys. \textbf{50}, 015102 (2023).
\bibitem{Soylu2012epja} A. Soylu \textit{et al.}, Eur. Phys. J. A \textbf{48}, 128 (2012).
\bibitem{Wei2017prc} K. Wei, H.F. Zhang, Phys. Rev. C \textbf{96}, 021601(R) (2017).

\bibitem{saxenanew2023} G. Saxena, A. Jain, P.K. Sharma, Prafulla Saxena, Communicated to Scientific Reports (2023).
\bibitem{dengplb2021} Jun-Gang Deng, Hong-Fei Zhang, Phys. Lett. B \textbf{816}, 136247 (2021).
\bibitem{Denisov2009prc} V.Yu. Denisov, A.A. Khudenko, Phys. Rev. C \textbf{79}, 054614 (2009).
\bibitem{huang1976} K.N. Huang \textit{et al.}, At. Data Nucl. Data Tables \textbf{18}, 243 (1976).
\bibitem{ws42014} N. Wang, M. Liu, X. Wu, J. Meng, Phys. Lett. B \textbf{734}, 215 (2014).
\bibitem{moller2019} P. M\"{o}ller, M.R. Mumpower, T. Kawano, W.D. Myers, At. Data Nucl. Data Tables \textbf{125}, 1 (2019).
\bibitem{hfb2004}  J. Dobaczewski, M.V. Stoitsov, W. Nazarewicz, AIP Conference Proceedings \textbf{726}, 51 (2004).
\bibitem{saxenaPLB2019} G. Saxena, M. Kumawat, M. Kaushik, S.K. Jain, Mamta Aggarwal, Phys. Lett. B \textbf{788}, 1 (2019).
\bibitem{saxena2017} G. Saxena, M. Kumawat, M. Kaushik, U.K. Singh, S.K. Jain, S. Somorendro Singh, M. Aggarwal, Int. J. Mod. Phys. E \textbf{26}, 1750072 (2017).
\bibitem{saxenaplb2017} G. Saxena, M. Kumawat, M. Kaushik, S.K. Jain, M. Aggarwal, Phys. Lett. B \textbf{775}, 126 (2017).
\bibitem{singh2020}     U.K. Singh, P.K. Sharma, M. Kaushik, S.K. Jain, Dashty T Akrawy, G. Saxena, Nucl. Phys. A \textbf{1004}, 122035 (2020).
\bibitem{Fiset1972} E.O. Fiset, J.R. Nix, Nucl. Phys. A \textbf{193}, 647 (1972).
\bibitem{Saxena2021jpg} G. Saxena, P.K. Sharma, P. Saxena, J. Phys. G: Nucl. Part. Phys. \textbf{48}, 055103 (2021).


\bibitem{rsharma2022} R. Sharma, A. Jain, P.K. Sharma, S.K. Jain, G. Saxena, Phys. Scr. \textbf{97}, 045307 (2022).
\bibitem{saxena2021QF} G. Saxena, A. Jain, P.K. Sharma, Phys. Scr. \textbf{96}, 125304 (2021).
\end{thebibliography}
\end{document}